\pdfoutput=1

\documentclass[twocolumn,a4paper,showpacs,amsmath,amssymb,aps,pre]{revtex4}

\usepackage{graphicx} % Include figure files
\usepackage{dcolumn} % Align table columns on decimal point
\usepackage{bm} % bold math
\usepackage{color}

\begin{document}

\preprint{draft}

\title{Two component dynamics of the superconducting order parameter revealed by time-resolved Raman scattering}

\author{R. P. Saichu$^{1}$, I. Mahns$^{1}$, A. Goos$^{1}$, S. Binder$^{1}$, P. May$^{1}$, S. G. Singer$^{1}$, B. Schulz$^{1}$, A. Rusydi$^{1,2}$, J. Unterhinninghofen$^{3}$, D. Manske$^{4}$, P. Guptasarma$^{5}$, M.S. Williamsen$^{5}$, and M. R\"{u}bhausen$^{1}$\footnote{corresponding author: ruebhausen@physnet.uni-hamburg.de}
}
%\thanks{corresponding author: ruebhausen@physnet.uni-hamburg.de}
%\email{ruebhausen@physnet.uni-hamburg.de}
\affiliation{$^{1}$Institut f\"{u}r Angewandte Physik, Universit\"{a}t Hamburg, Jungiusstrasse 11, D-20355 Hamburg, Germany. Center for Free Electron Laser Science (CFEL), Notkestrasse 85, D-22607 Hamburg, Germany\\
  $^{2}$Nanocore, Department of Physics, Faculty of Science, NUS, Singapore 117542, Singapore\\
  $^{3}$Institut f\"{u}r Theoretische Physik, Universit\"{a}t Bremen, Postfach 330440, D-28334 Bremen, Germany\\
  $^{4}$Max-Planck-Institut f\"{u}r Festk\"{o}rperforschung, Heisenbergstrasse 1, D-70569 Stuttgart, Germany\\
  $^{5}$Department of Physics, University of Wisconsin, Milwaukee, Wisconsin 53211, USA}

\date{\today}

\begin{abstract}

  We study the dynamics of the superconducting order parameter in the
  high-$T_c$ cuprate Bi$_2$Sr$_2$CaCu$_2$O$_{8-\delta}$ by employing a
  novel time-resolved pump-probe Raman experiment. We find two
  different coupling mechanisms that contribute equally to the pair
  breaking peak. One coupling sets in very fast at 2~ps and relaxes
  slow, while the other one is delayed and sets in roughly at 5~ps and
  relaxes fast. A model that couples holes through phonons is able to
  reproduce one part of the condensate dynamics, thus, we argue that
  hole-spin interactions are of importance as well.
\end{abstract}


\maketitle

The nature of the interaction between holes leading to superconductivity is encoded in the properties of the superconducting order parameter \cite{Scalapino, Fradkin}. These properties are reflected by the energy, the momentum dependence, and the time scales on which the order parameter reacts to an external perturbation \cite{Scalapino, Fradkin, Emery, Mihailovic}. In a material with competing interactions, there is a potential for the development of competing ordering phenomena \cite{Fradkin, Emery}. Undoped high temperature superconductors are antiferromagnetic insulators that become superconducting upon doping. In the superconducting state, the suppressed antiferromagnetic order might prevail on short length and time scales and, hence, affects thermodynamic properties \cite{Scalapino, Fradkin, Emery}. Therefore, it is crucial to understand the transient physics that is directly connected to a specific phase transition \cite{Mihailovic}. Such an approach allows the study of competing order parameters, their individual relaxation channels and elucidates the potential interplay between them.

Major progress in the development of pulsed laser, synchrotron, and free electron laser sources has led to innovative time-resolved techniques that can directly address the transient physics of, e.g. a correlated material~\cite{Kabanov, Kaindl1, Carr, Kaindl2, Demsar, Cavalleri}. However, several pump-probe techniques that have been applied to understand the superconducting condensate in high-temperature superconductors, fail to probe the superconducting order parameter $\Delta$({\textbf k}) directly \cite{Kaindl1}. In this context an intense debate exists on the question of whether or not one deals with one or more coupling processes for the pairing mechanism \cite{Demsar}. The two mechanisms which attracted the greatest attention invoke the coupling between holes through phonons or spin fluctuations \cite{Scalapino, Fradkin, Emery, Zeyer}. However, until today there is no clear understanding to which degree these mechanisms jointly contribute to the superconducting state. Different coupling mechanisms can be expected in different time scales in the response to an external perturbation. 

In this letter, we present a unique time-resolved two-color inelastic light (Raman) scattering experiment, which allows us to probe directly the superconducting order parameter. We report measurements of the dynamics of $\Delta$({\textbf k}). The samples are slightly overdoped Bi$_2$Sr$_2$CaCu$_2$O$_{8-\delta}$ (Bi-2212) high-temperature superconductors with a T$_c$ of 82~K and are very well characterized \cite{Budelmann, Rubhausen1, Klein2, Rubhausen2}. In our pump-probe Raman experiment we employ the UT-3 Raman spectrometer, which is a fully reflective achromatic spectrometer for the frequency range from deep ultraviolet to near infrared \cite{Schulz}. In order to obtain well defined time resolution we use two pulsed laser beams. The pump beam at 3.44~eV photon energy  [full width at half maximum (FWHM) is 0.8~ps] initiates the experiment and drives the system out of its equilibrium state. With the Raman probe beam at 1.72~ eV photon energy (FWHM $=$ 0.9~ps) we observe the energy and height of the pair breaking peak $\Delta$({\textbf k}) as a function of delay, i.e. time difference between the pump and the probe beam. The large energy difference between the pump and the probe beam avoids any spurious signal in the Raman probe. We have employed B$_{1g}$-polarization $\propto$ (x$^2$-y$^2$) resembling the {\it d\/}-wave symmetry of the superconducting order parameter. B$_{1g}$-symmetry can be studied by using crossed polarization between incident and scattered light with respect to the {\it a} and {\it b} axes in the CuO$_2$ planes. The pump beam heats the sample roughly 40-65~K above its equilibrium temperature, whereas the probe beam heats the sample by about 3~K. These values are estimated by comparing the non-equilibrium Raman spectra to the corresponding temperature corrected equilibrium Raman spectra from a continuous wave (CW) source, and numerical estimates \cite{Bock, Unterhinninghofen}. In the superconducting state the temperature rise is large enough to break Cooper pairs. Thus, changes in the energy and the height of the pair-breaking peak, i.e. the superconducting order parameter, can be probed.

\begin{figure}[t]
\includegraphics [width=8.0 cm]{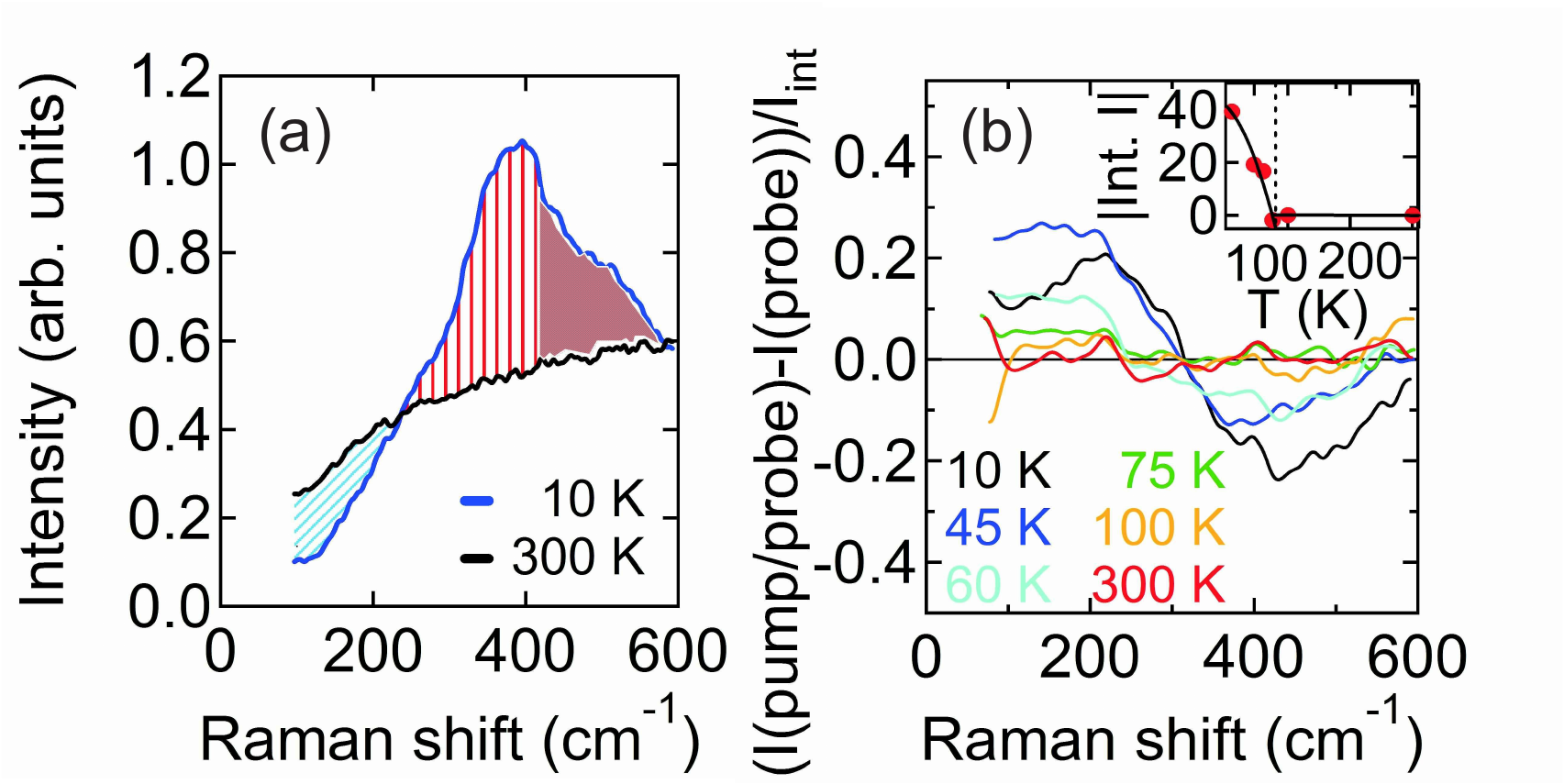}
\caption{(color online) Steady state and time-resolved Raman spectra of superconductors and metals depending on temperature in B$_{1g}$ geometry. (a) shows the spontaneous Raman scattering intensity at 10~K and at 300~K. A gap opens below 250~cm$^{-1}$ (blue area) and a pair breaking peak appears around 420~cm$^{-1}$ (red area). (b) Temperature evolution of the Raman difference spectrum at a delay between the pump and the probe beam of 3~ps. Positive values indicate an increase of intensity in the pumped state compared to the equilibrium state, negative values indicate the opposite. The inset shows the integrated intensity of the difference spectra between 300~cm$^{-1}$ and 600~cm$^{-1}$.}
\label{fig:Fig1}
\end{figure}

In Fig.~\ref{fig:Fig1}(a) we show Bose-corrected steady state spectra for a transferred energy (Raman shift) between 100~cm$^{-1}$ and 600~cm$^{-1}$ in the normal and superconducting state of Bi-2212. We employ a probe energy of 1.72~eV in order to minimize resonance effects that are known to occur for higher incident photon energies \cite{Budelmann, Deveraux}. In the normal state one can observe a flat background (black curve) that results from the scattering of charges within a marginal Fermi liquid \cite{Klein1, Hackl, Deveraux, Varma}. On the other hand, in the superconducting state a gap opens (blue shaded area) and a pair breaking peak forms at roughly twice the maximum value of $\Delta$({\textbf k}) (red shaded area) at $\approx$ 420~cm$^{-1}$ $\cong$ 52~meV \cite{Deveraux}. The height of the peak is proportional to the number of Cooper pairs that are broken around ($\pm\pi$,0) and (0,$\pm\pi$) in the Brillouin zone.

If we pump the superconducting state, we raise the sample temperature. For different equilibrium temperatures and at a fixed delay of 3~ps Raman difference spectra normalized to the integrated scattering intensity~$I_{int}$ are shown in Fig.~\ref{fig:Fig1}(b). In the pumped state, we clearly find a difference spectrum with positive values at low energies and negative values at higher energies. This observation reflects directly a loss of Cooper pairs and correspondingly a decreased intensity in the pair breaking peak.  Due to charge conservation, this decrease of intensity is accompanied by an increase of quasi-particle spectral weight within the gap. Finally, the Raman difference spectrum vanishes at T$_c$. The integral over the magnitude of the Raman difference spectrum is shown in the inset of Fig.~\ref{fig:Fig1}(b) clearly demonstrating that we observe effects that are related exclusively to the superconducting state and are not related to the pseudogap \cite{Emery, Mihailovic}.

\begin{figure}[t]
\includegraphics [width=8.6 cm]{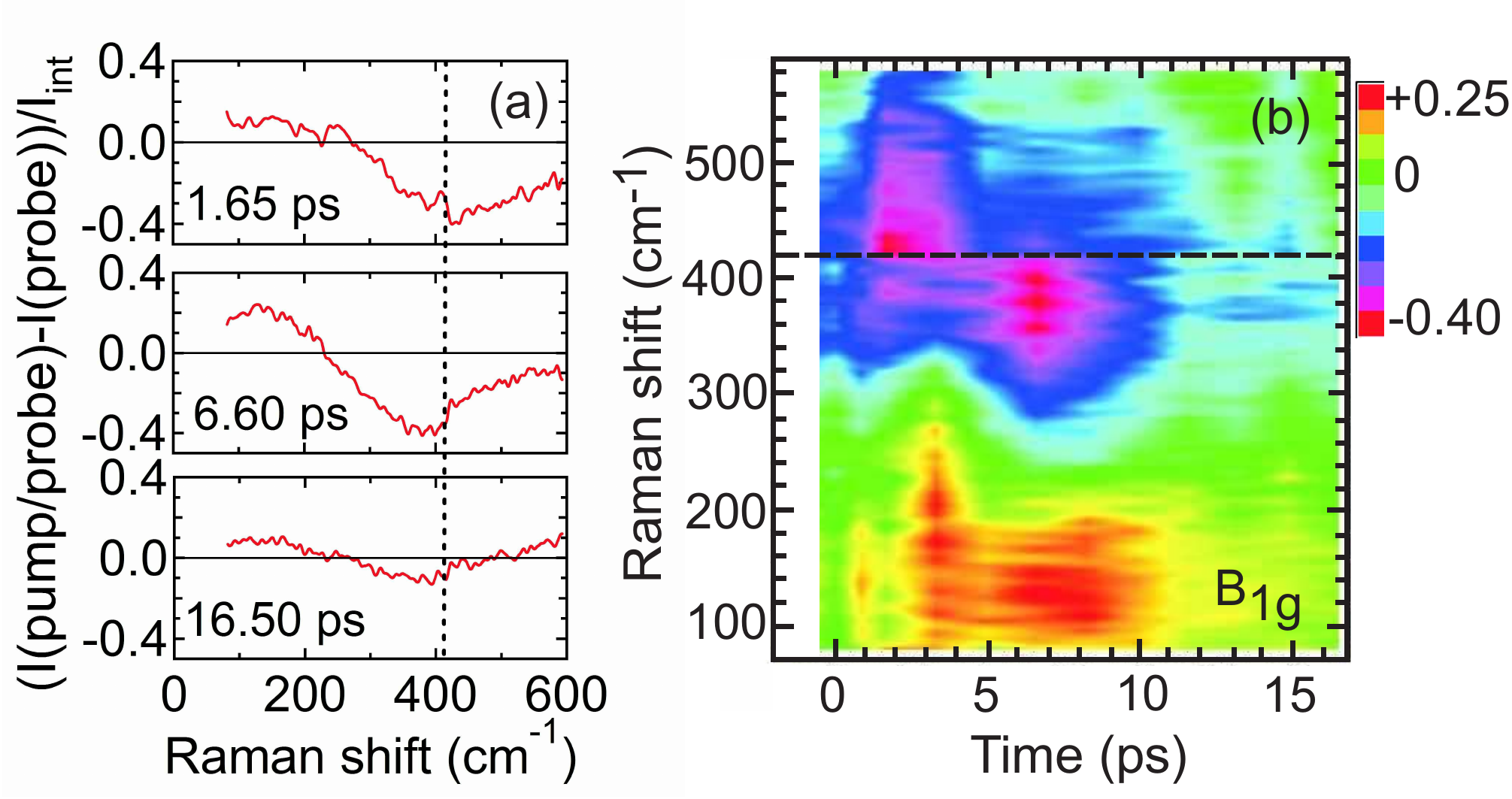}
\caption{(color online) Temporal evolution of the time-resolved Raman difference spectra at an equilibrium temperature of 10~K in B$_{1g}$ geometry. In (a) three difference spectra of three different delay times are shown. The contour plot shown in (b) presents 12 Raman difference spectra for different delay times. The dashed line separates two energy regions of the pair-breaking peak that reveal different characteristic behavior. The intensity changes are color coded demonstrating the transfer of spectral weight from high to low energies after 1~ps, respectively.}
\label{fig:Fig2}
\end{figure}

In order to analyze the dynamics of Cooper pairs in the pumped spectrum, we show of example in Fig.~\ref{fig:Fig2}(a) three Raman difference spectra at 10~K as a function of delay time between pump and probe pulse. Already at 1.65~ps after the pump beam, the Raman probe detects a pronounced loss of spectral weight at high energies which corresponds to the high-energy tail of the pair-breaking peak shown in Fig.~\ref{fig:Fig1}(a), and only a marginal gain of spectral weight within the gap. At 6.6~ps the loss of spectral weight is shifted to energies around the maximum intensity in the pair-breaking peak and a clear gain of spectral weight within the gap [blue shaded area in Fig.~\ref{fig:Fig1}(a)] is observed.  Furthermore, the integrated change at 6.6~ps is significantly larger as compared to the change at 1.65~ps. This clearly demonstrates that the superconducting condensate reacts on two different time scales to the pump beam.  Interestingly, these time scales are roughly connected to two sides of the pair-breaking peak, i.e. above and below 420~cm$^{-1}$. After 16.5~ps the difference spectrum is overall strongly diminished indicating a complete relaxation back into the equilibrium state.

In Fig.~\ref{fig:Fig2}(b) we show the temporal evolution in more detail by employing a density plot consisting of 12 Raman difference spectra (green color corresponds to no intensity change, i.e.  the equilibrium state). The Raman shift is set on the y-axis, whereas the time delay between pump and probe beams is set on the x-axis. From this data set we can derive the two different time scales in more detail: First, a fast response at around 2~ps that starts with a suppression of a pair breaking peak above 420~cm$^{-1}$. This time scale relates directly to the results of novel time-resolved ARPES experiments by Perfetti {\it et al.\/} showing unambiguously that the hot electrons and hot phonons created by a pump pulse thermalize within 50~fs and 2~ps, respectively \cite{Perfetti}. From this we can conclude that our observed dynamics after 2~ps is driven by a homogeneously heated sample. After the initial loss of spectral weight in the pair-breaking peak around 2~ps, an increase of low-energy spectral weight is observed roughly 1~ps later, reflecting the transformation of holes from the superconducting condensate into quasiparticles due to charge conservation. Then, a second, delayed response after 5~ps is observed in the pair breaking peak, this time below 420~cm$^{-1}$. This spectral weight suppression in the pair-breaking peak also yields an additional gain in spectral weight within the gap roughly after 1~ps. Thus, summarizing our spectra, we can clearly identify two contributions to the suppression of the pair-breaking peak that have their typical energy and time scales. We observe a fast onset of suppression of the pair-breaking peak above 420 cm$^{-1}$ and a delayed suppression below 420 cm$^{-1}$. Both regions of the pair-breaking peak are indicated by the differently red shaded areas in Fig.~\ref{fig:Fig1}(a). As expected from charge conservation, both responses yield their respective gain of spectral weight within the gap indicating a clear redistribution of the superconducting condensate [blue shaded area in Fig.~\ref{fig:Fig1}(a)].

\begin{figure}[t]
\includegraphics [width=7.8 cm]{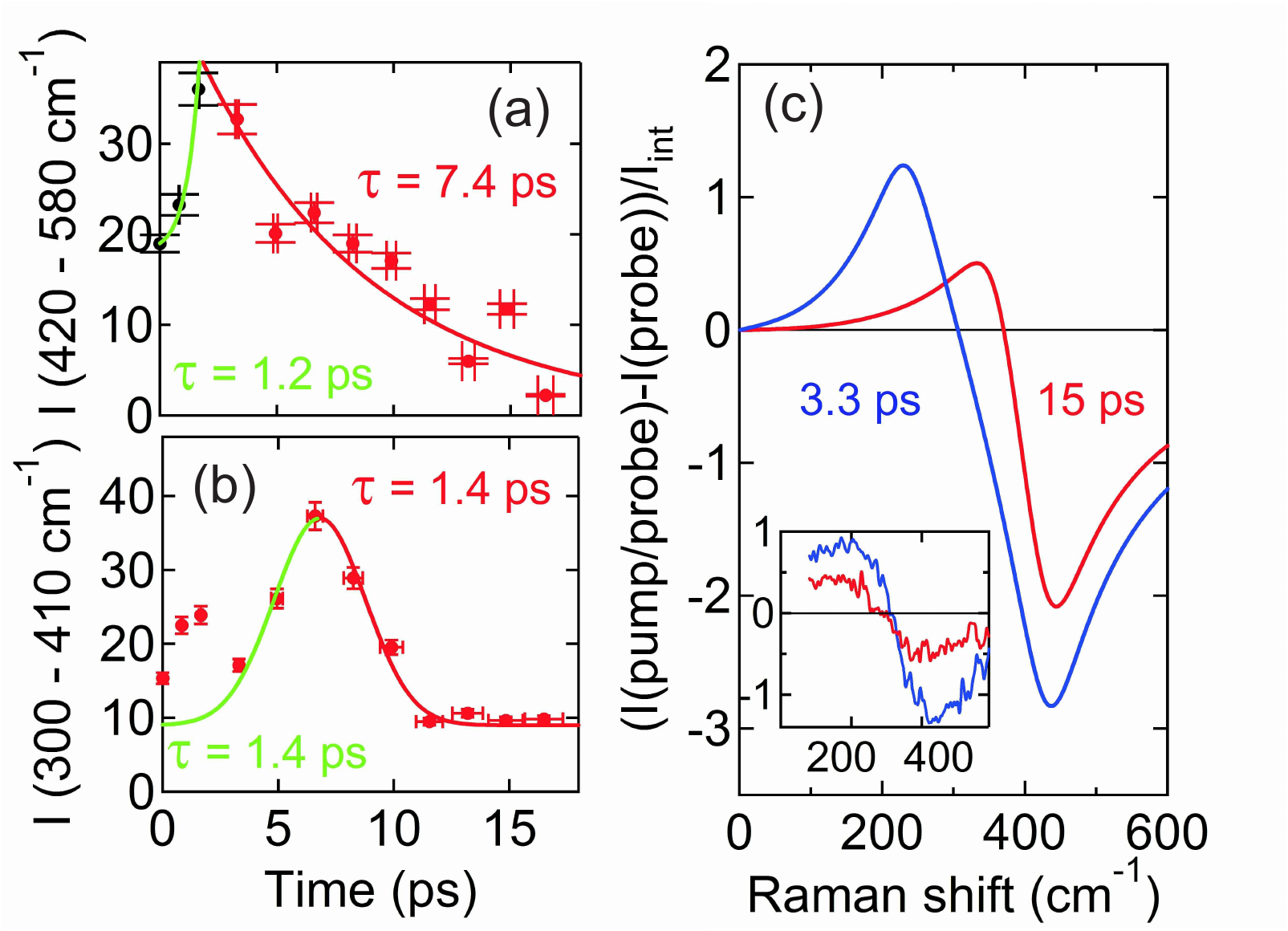}
\caption{(color online) Integrated intensity for the two different energy regions and comparison between theoretical and experimental response. (a) Energy region from 420~cm$^{-1}$ to 580~cm$^{-1}$. (b) Energy region from 300~cm$^{-1}$ to 410~cm$^{-1}$. The green characteristic decay time of (a) $\tau=$ 7.4~ps and (b) $\tau=$ 1.4~ps are determined by an exponential and Gaussian fit, respectively. (c) Calculated time-resolved Raman difference spectra for two different delay times. The inset shows the experimental difference spectra.}
\label{fig:Fig3}
\end{figure}

Having identified two different time scales of the superconducting condensate, what is their corresponding time (decay) constant? To study this, we have integrated the spectral weight along the energy axis above and below 420~cm$^{-1}$ as indicated by the dashed line in Fig.~\ref{fig:Fig2}(b). The temporal evolution of these spectral-weight changes are displayed in Fig.~\ref{fig:Fig3}(a) and (b), respectively. As we will argue below, it is remarkable that the behavior of the fast high energy response is quite consistent with a relaxation process through in-plane phonons. Our calculation will show that this characteristic time scale of the relaxation process of about 7.4~ps is consistent with previous estimates from other experiments \cite{Kaindl1, Demsar}. However, the second, delayed response below 420~cm$^{-1}$ is unexpected and has also an untypical behavior, since rise and decay times are roughly equal and of only 1.4~ps [see Fig.~\ref{fig:Fig3}(b)]. From neutron scattering experiments we know that for energies below 420~cm$^{-1}$ the coherence peak in the spin susceptibility yields to a strong coupling between holes even close to T$_c$ \cite{Capogna}. Thus, it is reasonable to assume that remnant spin-mediated coupling between holes yields an enhanced stiffness of the paired holes against an external perturbation resulting in this delayed reduction of the Cooper pairs after 5~ps. Such a scenario could be supported by a picture in which charge rich domains are surrounded by fluctuating charge poor domains. Indeed, static domains have been observed by static probes such as Scanning Tunneling Microscopy (STM) experiments~\cite{Slezak}.

Finally, we show in Fig.~\ref{fig:Fig3}(c) a comparison between a model and our Raman difference spectra in the pumped state. For our calculations, we have employed the method of density-matrix theory \cite{Richter, Kira} which has been recently generalized for time-resolved spectroscopy on superconductors. We have used the Hamiltonian of Ref.~\cite{Unterhinninghofen} that invokes the coupling to the most important phonon modes of the copper-oxygen planes, i.e. the breathing and buckling modes. The corresponding electron-phonon matrix elements are treated within LDA. Coupling of holes to spin fluctuations is not considered. Then, in the superconducting state, the non-equilibrium Raman intensity can be calculated from the imaginary part of the response function
\begin{eqnarray}\nonumber\label{1}
 \chi({\bf q}=0 , \, \omega) &\sim &  \sum\limits_{\textbf k} \gamma_{\textbf k}^2 
 \Big( u_{\textbf k}^2 \big \langle \alpha_{\textbf k}^{\dagger}
 \alpha_{\textbf k} \big \rangle + v_{\textbf k}^2 \big(1-\big
  \langle \beta_{\textbf k}^{\dagger} \beta_{\textbf k} \big \rangle \big) \\
  && + \, u_{\textbf k} v_{\textbf k} \big( \big\langle \alpha_{\textbf k}^{\dagger}
  \beta_{\textbf k}^{\dagger}\big\rangle + \big\langle \beta_{\textbf k}
  \alpha_{\textbf k}\big\rangle\big) \Big)\: .
\end{eqnarray}

\noindent ${\bf q}$ and $\omega$ denote the transferred momentum and energy, respectively. The summation runs over all (electronic) wavevector ${\bf k}$ in the first Brillouin zone. $\gamma_{\textbf{k}}$ $\propto (k_x^2 - k_y^2)$ is the non-resonant $B_{1g}$ Raman vertex. Due to the employed Bogoliubov transformation the Raman susceptibility $\chi({\bf q}=0 , \, \omega)$ includes the mixing amplitudes $u_{\textbf k}$, $v_{\textbf k}$ and the creation and annihilation operators $\alpha_{\textbf k}$, $\alpha_{\textbf k}^{\dagger}$, $\beta_{\textbf k}$ and $\beta_{\textbf k}^{\dagger}$ of the Bogoliubov quasiparticles (linear combination of holes and electrons), respectively \cite{note}. The use of the low photon energies in the probe beam of about 1.7~eV makes this non-resonant vertex more applicable as compared to visible and UV-photon energies where strong resonance effects need to be considered in Bi-2212 \cite{Budelmann}. We chose a one band tight-binding fit to the measured band structure \cite{Deveraux, Kordyuk} and a {\it d\/}-wave order parameter $\Delta_{\textbf k}$ = $\Delta _0$ (cos$k_x $- cos$k_y$)/2, yielding a quasiparticle dispersion $ E_{\textbf k} =\sqrt {(\epsilon_{\textbf k}^2 - \mu)^2 + \Delta_{\textbf k}^2}$. Then, we calculate the time-dependent expectation values $\big \langle \alpha_{\textbf   k}^{\dagger} \alpha_{\textbf k} \big \rangle(t)$, $ \big \langle \beta_{\textbf k}^{\dagger} \beta_{\textbf k} \big \rangle(t)$ and $ \big \langle \beta_{\textbf k} \alpha_{\textbf k} \big \rangle(t)= \big\langle\alpha_{\textbf k}^{\dagger} \beta_{\textbf k}^{\dagger} \big \rangle^{*} (t)$ with the help of coupled Boltzmann-equations for the non-equilibrium situation in the pumped state. An additional damping $\delta$ $=$ 5~meV is used to account for other scattering processes that we do not take into account. After numerical solution of the equations of motion, the results can be inserted into Eq.~(\ref{1}), and a pump-probe difference Raman spectrum as a function of delay time can be readily calculated.

Figure~\ref{fig:Fig3}(c) shows the calculated difference Raman spectra at a fixed delay time of 3.3~ps and 15~ps. The inset shows the corresponding measured data that have been corrected for the ratio of the pumped to the probed volume \cite{remark}. We find a fair agreement considering that we have used no adjustable fitting parameters in the normalized plot of the Raman response. However, the relaxation in the measured Raman spectra is faster as compared to the model calculations indicating a superposition of different phonon modes that contribute to the relaxation process. The electron-phonon coupling strength for the in-plane oxygen breathing and out-of-plane buckling modes \cite{Song} yields typical time constants of 4~ps and 20~ps, respectively (not shown). From this we derive a roughly 80\% contribution of the breathing mode in the phonon mediated relaxation process \cite{Unterhinninghofen, Richter}. Furthermore, it is obvious that the second response below 52~meV with a delayed onset after 5~ps and very fast decay time of less then 1.4~ps cannot be described within our model. As mentioned above, this strongly suggest that the hole-spin interaction and the inhomogeneous distribution of holes within the copper-oxygen plane needs to be included in any theory that aims to fully understand the time-resolved Raman response as shown in Fig.~\ref{fig:Fig2}(b).

In conclusion, we present a unique two-color Raman experiment revealing the ultrafast dynamics of the superconducting order parameter in Bi-2212 by employing a novel time-resolved pump-probe Raman experiment. Our results clearly demonstrate that the pair-breaking peak in the Raman responses reacts on two different time scales. These time scales are equivalent to two different coupling mechanisms. Both couplings show the redistribution of spectral weight from the pair-breaking peak to a quasiparticle response within the gap. The first, fast response sets in at 2~ps and relaxes within 7.4~ps, the second response sets in at 5~ps and relaxes within 1.4~ps, respectively [see Fig.~\ref{fig:Fig2}(b)]. Based on density-matrix calculations, we are able to model the first response by the coupling of holes to phonons and find the in-plane breathing mode to be dominant. However, for an understanding of the second, delayed Raman response one would need to account explicitly for hole-spin interactions and for the inhomogeneous nature of the charge distribution. Our new technique provides direct information on the condensate dynamics as well as on the coupling mechanisms responsible for the formation of the superconducting state.

We acknowledge stimulating discussions with S. L. Cooper, M. V. Klein, T. P. Devereaux, J. Schneider, M. Drescher, W. Wurth, K. Scharnberg, G. Sawatzky, and A. Cavalleri. Financial support has been granted by the DFG via Ru773/3-1, GRK 1286, and the BMBF via the project "Time Resolved VUV Raman Scattering".

\bibliographystyle{plain}
\bibliography{test}

\begin{thebibliography}{28}
\bibitem{note}
One can recover the equilibrium Raman response from Eq. (\ref{1}) by calculating the equilibrium expectation values; one finds $\big \langle \alpha_{\textbf k}^{\dagger} \alpha_{\textbf k} \big \rangle =  \big \langle \beta_{\textbf k}^{\dagger} \beta_{\textbf k} \big \rangle = f(E_{\textbf k})$ with the Fermi function~{\it f\/}, $\big \langle \beta_{\textbf k} \alpha_{\textbf k} \big \rangle = u_{\textbf k} v_{\textbf k}(1-2f(E_{\textbf k}))/(\hbar \omega +i\eta +2E_{\textbf k})$, which yields the usual Raman response.
\bibitem{remark} Both volumes are not identical, since the penetration depths $\lambda$ are different. The difference is due to the variation of the dielectric function between 1.72 eV ($\lambda$~=~120~nm) and 3.44 eV ($\lambda$~=~50~nm).
\end{thebibliography}

\end{document}